\documentclass[prd,onecolumn,amsfonts,superscriptaddress,notitlepage]{revtex4-1}

\usepackage{amsmath}

\pdfoutput=1

\begin{document}


\title{AdS/CFT and the geometry of an energy gap}

\author{Andrew Hickling}
\affiliation{Theoretical Physics Group, Blackett Laboratory, Imperial College, London SW7 2AZ, UK } 
\author{Toby Wiseman}
\affiliation{Theoretical Physics Group, Blackett Laboratory, Imperial College, London SW7 2AZ, UK }


\date{June 2015}

\begin{abstract}
We consider a CFT defined on a static metric that is the product of time with a smooth closed space of positive scalar curvature. 
We expect the theory to exhibit an energy gap and our aim is to investigate how that gap depends on the geometry of the space.
For a free conformal scalar it is straightforward to show the gap normalised by the minimum value of the Ricci scalar of the space is minimised when the space is a sphere.
Our main result is then to show using geometric arguments that precisely the same result holds for fluctuations of a scalar operator in any holographic CFT. 
We prove this under the assumption that the dual vacuum geometry is a smooth Einstein metric ending only on the conformal boundary, and then consider fluctuations of a minimally coupled massive scalar field about this.
We also argue the holographic CFT will have states dual to small bulk black holes whose existence is related to the energy gap.
We show the thermodynamic properties of these black holes obey a bound of a similar nature
to that of the scalar fluctuations - the ratio of CFT energy to a power of entropy for the states dual to black holes in the `small' limit is bounded from below when appropriately normalised by the minimum Ricci scalar of the boundary space. Again the bound is saturated for a sphere.

\end{abstract}

\maketitle

%
\section{Introduction}
%

One of the beautiful aspects of the AdS-CFT correspondence \cite{Maldacena:1997re} is that it provides a geometric description of various field theory phenomena in a strongly coupled context. For any $D$-dimensional CFT that has a holographic dual description a strongly coupled sector of the CFT dynamics decouples and is described by $(D+1)$-dimensional gravity with a negative cosmological constant.
This `universal gravity sector' is still very rich, with the degrees of freedom being given by bulk gravitons and black holes. Bulk spacetimes have prescribed asymptotics which are locally those of AdS. However, the conformal boundary does not have to be conformal to Minkowski. In principle we may choose any conformal boundary we wish, and this boundary geometry specifies the conformal class of the spacetime the CFT is defined on. Given this boundary data one must solve the bulk Einstein equation to find the dynamics of the universal sector of the CFT.
It is worth emphasising that while QFT techniques, such as the lattice, are able to extract information even for strongly coupled theories, it is very unclear how to practically use such techniques when the spacetime the QFT is defined on is not a simple one such as Minkowski. Thus AdS-CFT provides a very powerful tool to explore strongly coupled CFT on curved spacetimes. For a review of this see \cite{Marolf:2013ioa}.

We will focus on the case that the CFT is placed on a static spacetime so that the resulting theory has an energy gap. The canonical example is a CFT on a product of time with a unit sphere, so $\mathcal{M} = \mathbb{R}_t \times S^{D-1}$ where $S^{D-1}$ is a round unit $(D-1)$-sphere. This spacetime arises in the `radial quantisation' of a CFT and is simply related to Minkowski by a singular conformal transformation which maps the hamiltonian on $\mathcal{M}$ to the dilatation operator in the Minkowski conformal frame. The spectrum of energy eigenstates associated to the dynamics of a primary field is then determined algebraically, with the lowest eigenvalue given by the operator scaling dimension, and a discrete spectrum of excitations above this.

Consider now the CFT on a general static spacetime $\mathcal{M} = \mathbb{R}_t \times \Sigma$ where $\Sigma$ is a $(D-1)$-dimensional smooth closed space. We will generally consider the case that $\Sigma$ has positive scalar curvature, although some of our results will be more general. We will consider the case where $D>2$, so that $\Sigma$ has non-trivial geometry. 
In analogy with the sphere, we expect the CFT will exhibit an energy gap, with the scale of the gap being determined by the characteristic 
scale of the space $\Sigma$. Now however the spectrum cannot be deduced simply from a global conformal symmetry as this will generally be broken by $\Sigma$. The purpose of this paper is to explore the properties of such theories, and to characterise how this gap depends on the geometry of $\Sigma$. Since this problem has a geometric character we will find that AdS-CFT allows powerful statements to be made for strongly coupled CFTs with a gravity dual.

The gap is not a scale invariant quantity. Scaling the size of the space $\Sigma$ simply scales the gap. In order to eliminate this trivial dependence on $\Sigma$, we will consider how a scale invariant `normalised gap' depends on the geometry of $\Sigma$, where this normalised gap is constructed by dividing the gap by some geometric invariant of $\Sigma$. One might for example take an appropriate power of the volume. However, one lesson we will learn is that a natural invariant appears to be the minimum value of the Ricci scalar of $\Sigma$. 
Note that since we will largely consider positive scalar curvature spaces this is a good quantity to normalise by.
Our starting point is the observation that for a free conformal scalar in $D$-dimensions it is simple to show that the gap normalised in this manner is bounded from below, with the bound being saturated when $\Sigma$ is a sphere. The question is then whether such a simple result can extend to a strongly coupled setting, and we use the AdS/CFT correspondence to investigate this.
In the case of a holographic CFT, the vacuum of the canonical case of $\mathcal{M} = \mathbb{R}_t \times S^{D-1}$ is described by a dual global AdS spacetime. As discussed in \cite{Witten:1998zw,Sundborg:1999ue,Aharony:2003sx} fluctuations about this exhibit a gap, and have the interpretation of glueball excitations about a confining vacuum in the case the CFT is a gauge theory. 
The spectrum of fluctuations for a bulk field is determined by the scaling dimension of the dual operator to that field.
However, in addition to small fluctuations about the vacuum, the universal gravity sector also contains black holes which describe states of the deconfined thermal phase.
These exhibit a gap in the sense that the there is a minimum temperature for static black holes asymptotic to global AdS, with this minimum temperature being determined by the size of the boundary sphere. Such black holes split into two branches, the `small' and the `large', and there is the Hawking-Page phase transition between global AdS and the `large' branch of black holes \cite{Hawking:1982dh}. In the CFT this is interpreted as a first order phase transition between the confined and deconfined phases. 
If we now consider deforming the conformal boundary metric to $\mathcal{M} = \mathbb{R}_t \times \Sigma$, then provided one can solve the static bulk Einstein equations the vacuum of the holographic CFT can again be explored. 
We require that this bulk static vacuum metric only ends on the usual conformal boundary, and away from that is smooth. 
In particular we require there to be no extremal horizons, which would generically not support a gapped spectrum (as for example in the case of Poincare-AdS). 

It is reasonable to expect such Einstein metric duals to exist for generally classes of smooth closed $\Sigma$. For $D = 3$ existence has been proven in the case that $\Sigma$ has positive scalar curvature \cite{Anderson:2001pf,Anderson:2002xb}, and hence by Gauss-Bonnet $\Sigma$ 
is of sphere topology. 
Given such a bulk exists, we expect a gapped spectrum for fluctuations of bulk fields about it, and analogous black hole behaviour to that of global AdS.
A well known example of this (although with vanishing scalar curvature) is for $\Sigma = T^{D-2} \times S^1$, with $T^{D-2}$ a torus, and on the circle Scherk-Schwarz boundary conditions are taken. Then the dual geometry is the AdS-soliton \cite{Horowitz:1998ha}, which is a  prototype for phenomenological models of confinement  \cite{Witten:1998zw,Sakai:2004cn}. The spectrum of fluctuations is again gapped, and an analogous black hole behaviour to global AdS is thought to occur  \cite{Aharony:2005bm,Horowitz:2007fe,Emparan:2009dj,Figueras:2014lka}. Another example in $D=4$ is that for $\Sigma = S^2 \times S^1$ with the same antiperiodic fermion boundary conditions on the circle which has been constructed numerically \cite{Copsey:2006br}.

Our first result is to show that provided a dual static vacuum geometry exists satisfying our conditions (smoothness and ending only on the conformal boundary) then the spectrum of minimally coupled scalar fluctuations about this is discrete, and just as for the free conformal scalar, we prove a lower bound on the normalised gap, where again this is normalised by the minimum value of the Ricci scalar of $\Sigma$. As in the free case, this bound is saturated when $\Sigma$ is a sphere. 
It is worth emphasising that in a generic top-down holographic model (for example \cite{Skenderis:2006uy,Gauntlett:2009zw}) with a consistent truncation to the $(D+1)$-dimensional universal gravity sector, then a fluctuation mode in the full model about a solution to this gravity sector which behaves as a scalar in $(D+1)$-dimensions will appear as a minimally coupled massive scalar.\footnote{Note we have assumed the top-down model is second order in derivatives, and then the only curvature couplings to a scalar fluctuations can be via the $(D+1)$-dimensional Ricci scalar which is constant and hence simply contributed to the mass.}
Hence considering fluctuations in a minimal bulk scalar is not a restriction, but in this context is the generic dual to a CFT scalar operator with a gravitational description.
This bound is surprising in the sense that it is precisely the same for both the free and holographic (and hence strongly coupled) CFTs. In particular, the normalised gap entering the bound, which measures the gap with respect to the minimum Ricci scalar of $\Sigma$, emerges naturally in both cases, although in quite different ways. This motivates the question of whether such a bound applies more widely to general CFTs.

We then consider the holographic CFT at finite temperature. We argue that in general small and large static black holes analogous to those in global AdS exist. The behaviour of very large black holes is universal, and does not depend on the details of $\Sigma$. However, the behaviour of very small black holes is intimately related to the fact there is a gap, and depends on the detail of $\Sigma$.
Our second main result is to prove a bound on the thermodynamics of the very small black holes. Interestingly this bound again is given in terms of the minimum value of the Ricci scalar of $\Sigma$, and is saturated for the case that $\Sigma$ is a sphere. 

The structure of the paper is as follows. We begin in section~\ref{sec:confscalar} considering the bound obtained on the lowest energy excitation for a free conformally coupled scalar field on the static spacetime, $\mathcal{M} = \mathbb{R}_t \times \Sigma$ with $\Sigma$ closed.
Then in section~\ref{sec:holo} we review the universal dual gravity sector of a strongly coupled holographic CFT. We discuss the bulk geometry that describes the vacuum for the CFT on $\mathcal{M}$, and also fluctuations of a bulk scalar about this geometry which describe dynamics dual to a scalar primary operator in the CFT. We also introduce our main technical tool which is a minimum principle for a particular curvature scalar in the bulk. In section~\ref{sec:holoscalar} we derive a lower bound on the energy of a bulk scalar excitation that is precisely of the same form as for the free conformal scalar. In section~\ref{sec:holobh} we go on to discuss the behaviour of static bulk black holes, and in particular describe the properties of very small black holes, and a bound on these properties which involves the same geometric data of $\Sigma$ as the scalar fluctuation bound.
We end with a discussion in section~\ref{sec:discussion}.

%
\section{A free conformal scalar}
\label{sec:confscalar}
%

Consider a free real conformally coupled scalar on a $D$-dimensional spacetime with metric $g^{(D)}$. 
This obeys the wave equation
\begin{equation}
\nabla_{(D)}^2 \psi = \frac{D-2}{4(D-1)} {R^{(D)}} \psi
\end{equation}
with conformal coupling to the metric via its Ricci scalar ${R^{(D)}}$. This theory is a CFT, with the scalar field $\psi$ being a primary with scaling dimension $\Delta = (D-2)/2$.
Now place this CFT on the static  spacetime $\mathbb{R}_t \times \Sigma$, with metric
\begin{eqnarray}
\label{eq:CFTspacetime}
ds^2 = -dt^2 + \bar{g}_{ab}(x) dx^a dx^b
\end{eqnarray}
where $a,b = 1, \ldots, D-1$ so that $(\Sigma, \bar{g})$ is a $(D-1)$-dimensional Riemannian manifold. 
Let us assume the spatial geometry $\Sigma$ is smooth and closed, so compact and without boundary. 
One might think we have made a choice in specialising to a static metric where the lapse function is trivial - an `ultra-static' metric. However, due to the conformal invariance, in fact we have only made a restriction that the vector field $\partial / \partial t$ is globally timelike, so that we may select a conformal frame where the metric takes the above form.

We now consider the spectrum of this free theory. Let us write a general solution in terms of energy eigenstates, so $\psi = \sum_{\{\omega\}} f_\omega e^{i \omega t}$.
Since $\Sigma$ is closed, the energy of a mode is derived from the eigenvalues of a self-adjoint elliptic operator on $\Sigma$, 
\begin{equation}
\left[ - \bar{\nabla}^2 + \frac{D-2}{4(D-1)} \bar{R} \right] f_\omega = \omega^2 f_\omega 
\end{equation}
where $\bar{\nabla}$ and $\bar{R}$ are the covariant derivative and Ricci scalar of $(\Sigma,\bar{g})$.
Hence the spectrum of the eigenvalues, $\omega^2$, 
will be real and discrete.
Let us assume we are interested in stable theories, and so $\omega_{}^2 \ge 0$. We define the energy gap, $\omega_{min}$, as the energy of the lowest mode, so $\omega_{min} = \min_{\{\omega\}}  |\omega |$.
A natural question to ask is how the spatial geometry affects the energy gap.
The fact that $\Sigma$ is closed allows us to integrate by parts on $\Sigma$ to deduce,
\begin{eqnarray}
 \omega^2 \int_\Sigma \sqrt{\bar{g}} \, f^2_\omega = \int_\Sigma \sqrt{\bar{g}} \left(  \frac{D-2}{4(D-1)} \bar{R} \, f^2_\omega + \left({\partial} f_\omega \right)^2 \right)
\end{eqnarray}
and hence gives the following bound,
\begin{eqnarray}
\label{eq:freebound}
\omega^2 \ge \frac{D-2}{4(D-1)} \bar{R}_{min}
\end{eqnarray}
where $\bar{R}_{min}$ is the minimum value of the Ricci scalar of $\Sigma$. 
If $\bar{R}_{min} < 0$ our bound gives no information provided we already know the CFT is stable. 
Hence we now restrict $\Sigma$ to have positive scalar curvature so that $\bar{R}_{min} > 0$, and then the result above provides a bound on $\omega_{min}$, the energy gap of the theory.
We have equality if and only if $\bar{R}$ and $f_\omega$ are constant. 

The observation that a sphere saturates this bound (since $\bar{R}$ is constant and
the eigenmodes $f_\omega$ are given by spherical harmonics, with the lowest mode being constant)
motivates rewriting it as,
\begin{eqnarray}
\label{eq:freebound2}
\Omega \equiv {\omega_{min}}{\mathcal{R}}  \ge \Delta \; , \quad \bar{R}_{min} = \frac{(D-1)(D-2)}{\mathcal{R}^2}
\end{eqnarray}
where $\mathcal{R}$ is 
the radius of a sphere whose scalar curvature is given by $\bar{R}_{min}$. We have introduced the normalised energy gap, $\Omega \equiv \omega_{min}  \mathcal{R}$. While the energy gap $\omega_{min}$ has mass dimension, $\Omega$ is dimensionless and hence scale invariant. In this form, the bound encodes how the gap depends on the geometry of $\Sigma$ modulo trivial scaling. 

Is it surprising that for a sphere the bound saturates to a simple form $\Omega = \Delta$? It is indeed non-trivial that a sphere saturates the bound. However, the simple form of $\Omega_{}$ in the case that $\Sigma$ is a sphere is precisely to be expected. This is due to the fact  that in the case of a unit sphere the spectrum of energies is precisely determined by a residual global conformal symmetry, since $\mathbb{R}_t \times S^{(D-1)}$ is conformally related to Minkowski spacetime, being the conformal frame used for radial quantisation of a CFT. The dilatation operator in the Minkowski frame is mapped to the Hamiltonian for $\mathbb{R}_t \times S^{(D-1)}$ with a unit sphere. 
Given a scalar primary field with scaling dimension $\Delta$, there is a corresponding energy eigenstate on $\mathbb{R}_t \times S^{(D-1)}$ with energy given by $\Delta$ that is the lowest energy state associated to dynamics of this scalar, so that the one point function of the scalar doesn't vanish. 
Hence for the unit sphere, the lowest energy state associated to the dynamics of the scalar has energy $\omega_{min} = \Delta$. By scale symmetry, for a general radius sphere we obtain $\Omega = \Delta$.
Thus we should interpret the bound as saying the normalised gap, $\Omega_{}$, is minimised for $\Sigma$ being a sphere, when the minimum value is determined by conformal symmetry. To say it another way, knowledge that $\Omega$ is minimised for a sphere determines the right-hand side of the bound $\Omega \ge \Delta$, with the exact value of this right hand side constant adding no new information.
Note that if there exist multiple scalar primaries with distinct scaling dimensions, then we would expect a bound for each. These would then govern the lowest energy state associated to the dynamics of an individual scalar, with the one point functions of the other scalars vanishing.

It is interesting to note that if we did not consider a conformal scalar, but merely a minimal scalar, $\nabla^2_{(D)} \psi = 0$, then we would not have such a statement in terms of $\bar{R}_{min}$. The Lichnerowicz bound would bound the energy gap in terms of the norm of the Ricci tensor. However, with conformal coupling, it is the minimum value of the Ricci scalar that appears to control the behaviour of the gap. Interestingly we will see precisely the same structure emerge when we now consider a strongly interacting CFT using the AdS-CFT correspondence. While the origin of the dependence on $\bar{R}_{min}$ is straightforward to see in the free case, it will be far from obvious in the holographic context.

%
\section{A holographic CFT}
\label{sec:holo}
%

Let us consider the universal gravity sector of a holographic correspondence, where a $D$-dimensional CFT is dual to a $(D+1)$-dimensional bulk spacetime, with metric $g^{(D+1)}$, satisfying the Einstein condition,
\begin{eqnarray}
R^{(D+1)}_{\mu\nu} = - \frac{D}{\ell^2} g^{(D+1)}_{\mu\nu}
\end{eqnarray}
with $\ell$ giving the AdS scale, and related to the CFT central charge $c$ as, $c = \ell^{D-1} / 16 \pi G_{(D+1)}$, for bulk gravitational constant $G_{(D+1)}$.
We assume the bulk spacetime is dual to the CFT in its vacuum state on the static spacetime, $\mathbb{R}_t \times \Sigma$, with metric built from $(\Sigma, \bar{g})$ as,
\begin{eqnarray}
\label{eq:CFTmetric}
ds^2_{CFT} = -dt^2 + \bar{g}_{ab}(x) dx^a dx^b
\end{eqnarray}
where $a,b = 1, \ldots, D-1$, as for our free field example in equation \eqref{eq:CFTspacetime}. As in our previous example we take $\Sigma$ to be smooth and closed with the expectation that the CFT will exhibit an energy gap. 
We assume the bulk metric is static since we are interested in the dual to the vacuum state. We further assume the following;
\begin{enumerate}
\item The bulk is smooth away from the conformal boundary, with no other asymptotic regions or boundaries.
\item There are no bulk horizons and hence there is a globally timelike Killing vector.
\end{enumerate}
Given these assumptions we may build the static bulk spacetime from a $D$-dimensional Riemannian manifold $(\mathcal{M}, g)$ as,
\begin{eqnarray}
\label{eq:bulkmetric}
ds^2_{(D+1)} = g^{(D+1)}_{\mu\nu} dx^\mu dx^\nu = \frac{\ell^2}{Z^2} \left( -dt^2 + g_{ij} dx^i dx^j \right)
\end{eqnarray}
where $x^i$ are coordinates on $\mathcal{M}$,
with $i=1,\ldots D$ and $Z = Z(x^k) \ge 0$ is a function on $\mathcal{M}$. We take $\mathcal{M}$ to have a  boundary $\Sigma = \partial \mathcal{M}$, and $Z$ to be the defining function so that it vanishes only on the boundary of $\mathcal{M}$ with $d Z \ne 0$ there, which yields the locally AdS conformal boundary in the full spacetime. Then the conformal class of the boundary is given by equation \eqref{eq:CFTmetric} with $\bar{g}_{ab}$ being the metric induced from $g_{ij}$ on the boundary of $\mathcal{M}$. 
Given our assumptions we take $(\mathcal{M},g)$ to be smooth in its interior, and the function $Z$ to be smooth and positive over the interior of $\mathcal{M}$, which ensures $\partial/\partial t$ is globally timelike.
Since $Z$ is smooth and the bulk should have no other asymptotic regions, this implies that $\mathcal{M}$ has no asymptotic regions, and only the boundary $\Sigma = \partial \mathcal{M}$. In particular this implies that $\mathcal{M}$ has finite volume.

The metric $g_{ij}$ is known as the bulk \emph{optical} metric, as bulk null geodesics follow geodesic trajectories when projected into the Riemannian $(\mathcal{M}, g)$ \cite{OptMetric}. 
The function $Z$ has a physical interpretation as determining the local redshift in the bulk. More precisely we define the redshift function on $\mathcal{M}$, 
\begin{eqnarray}
{\xi}(x) = \frac{Z(x)}{\ell}
\end{eqnarray}
which relates the proper time $\tau$ of a static observer at some point $p \in \mathcal{M}$ to coordinate time $t$, so $dt = \xi_p d\tau$. 
Our choice of conformal frame is convenient as the boundary time translation Killing vector $\partial / \partial t$ is simply the restriction of the bulk vector $\partial / \partial t$ to the boundary. In particular this means that the redshift function $\xi(x)$ provides the redshift of the proper time of a static particle in the bulk at location $x$ relative to the boundary CFT time $t$.

The bulk static Einstein equations then can be reduced to $D$-dimensional equations over $\mathcal{M}$ as,
\begin{eqnarray}
\label{eq:bulkeqns}
R_{ij} &=& - \frac{(D-1)}{Z} \nabla_i \partial_j Z \nonumber \\
  R &=& \frac{D(D-1)}{Z^2} \left( 1 - \left( \partial Z \right)^2 \right)
\end{eqnarray}
where $R_{ij}$ and $R$ are the Ricci tensor and scalar of the optical metric $g_{ij}$, $\nabla$ is its covariant derivative, and indices are raised/lowered using this optical metric. These equations also imply $R = - \frac{(D-1)}{Z} {\nabla}^2 Z$.

We will be interested in considering time dependent small fluctuations of other bulk fields about a static solution to the above universal gravity sector. Here we will restrict to deformations that in $(D+1)$-dimensions are a real scalar field - dual to a scalar CFT operator. We are interested in small fluctuations so that this scalar equation is linear and we ignore its back reaction. In a generic top down construction there will be infinitely many such modes, some arising from scalar fields in the original supergravity, such as the dilaton, and others arising as Kaluza-Klein modes in a reduction on an internal space. 
Provided the original top down model is a usual two derivative gravitational theory, so that  the $(D+1)$-dimensional universal sector Einstein metrics lift to full solutions, then fluctuations of this scalar type will be described by a massive \emph{minimally coupled} bulk scalar since at the level of two derivatives and linear fluctuations the only covariant coupling of the scalar to the $(D+1)$-dimensional curvature is through the Ricci scalar, $R^{(D)}$, and this is constant for Einstein metrics and can be absorbed into the mass.
Thus we consider a real minimally coupled bulk scalar field $\phi$, with mass $m$, so, 
\begin{eqnarray}
\nabla^2_{(D+1)} \phi = m^2 \phi
\end{eqnarray}
and the only remnant of the top down origin of this fluctuation mode is the particular value of the mass $m$ which would have to be computed for each such scalar mode in a given top down model. 
The scaling dimension $\Delta$ of the dual scalar operator in the CFT is related to the mass as $m^2 \ell^2 = \Delta \left( \Delta - D \right)$ \cite{Gubser:1998bc,Witten:1998qj,Skenderis:2002wp}.
For this system to be well posed (and $\Delta$ to be real) we require the Breitenlohner-Freedman bound to be satisfied, namely $m^2 \ell^2 \ge - D^2/4$ \cite{Breitenlohner:1982jf,Breitenlohner:1982bm}.
We will only consider the intrinsic fluctuations associated to the scalar operator, and will not add external sources for it. Hence in the bulk we consider the usual `Dirichlet' boundary condition for $\phi$, namely that the leading behaviour near the conformal boundary is fixed to vanish leaving the subleading behaviour free. This implies we have also made the restriction that $\Delta > D/2$ so that the leading fall off is associated to the dual source \footnote{We note that we expect our arguments to generalise to the case of scalar fluctuations with $\Delta \le D/2$ but will not investigate this here.}. Note we then automatically satisfy the unitarity bound $\Delta \ge (D-2)/2$.

Since the bulk is static we will be interested in the spectrum of this linear scalar.
We perform a harmonic decomposition of the time dependence so that we may write a general solution as $\phi = \sum_{\{\omega\}} f_\omega e^{i \omega t}$,  
with the modes obeying,
\begin{eqnarray}
\label{eq:modeeqn}
- Z^{D-1} \nabla^i \left( \frac{1}{Z^{D-1}} \partial_i f_\omega \right) + \frac{\ell^2 m^2}{Z^2} f_\omega =  \omega^2  f_\omega
\end{eqnarray}
and as usual we may think of this as an elliptic eigenvalue problem on the optical geometry $(\mathcal{M},g)$, with $|\omega|$ giving the energy of the mode. We will assume that the CFT vacuum is stable to these scalar fluctuations so that $\omega^2 \ge 0$.

%
\subsection{Conformal Boundary}
%

Let us now examine the behaviour near the conformal boundary. We may choose a normal coordinate system to the boundary $Z = 0$ in the optical metric, so,
\begin{eqnarray}
ds^2_{(D+1)} = \frac{\ell^2}{Z^2(z, x^c)} \left( -dt^2 + dz^2 + g_{ab}(z, x^c) dx^a dx^b \right)
\end{eqnarray}
where $x^i = (z, x^a)$, and $z = 0$ corresponds to the conformal boundary. Now $z$ is interpreted as a bulk radial coordinate, and $x^a$ are spatial coordinates shared with the CFT.
Note that while these coordinates are not of Fefferman-Graham form, they do lead to an analogous series expansion in $z$ near the boundary,
\begin{eqnarray}
\label{eq:asym}
Z(z, x^c) & = & z \left( 1 - \frac{1}{6(D-1)(D-2)} \bar{R} z^2 + O(z^3) \right) \nonumber \\
g_{ab}(z, x^c) & = & \bar{g}_{ab}(x^c) - \frac{z^2}{D-2} \bar{R}_{ab} + O(z^3)
\end{eqnarray}
where $\bar{g}_{ab}(x^c)$ gives the spatial boundary metric in equation~\eqref{eq:CFTmetric}, and $\bar{R}_{ab}$ and $\bar{R}$ are its Ricci tensor and scalar. 
Note that, as usual, $(\partial_i Z)^2 = 1$ at $Z = 0$.
As for the Fefferman-Graham expansion, the higher terms in the expansion in $z$ are determined in terms of this boundary metric together with the boundary stress tensor, whose data appears at order $O(z^D)$ in the expansion.
From the expansion above we can   compute that the non-vanishing components of $R_{ij}$, have the values at the boundary,
\begin{eqnarray}
\left. R_{zz} \right|_{Z=0} = \frac{1}{D-2} \bar{R} \; , \quad \left. R_{ab} \right|_{Z=0}  = \frac{D-1}{D-2} \bar{R}_{ab}
\end{eqnarray}
and thus in particular the bulk optical Ricci scalar at the boundary, $\left. R \right|_{Z=0}$, is related to the boundary  Ricci scalar $\bar{R}$, as,
\begin{eqnarray}
\label{eq:Rbb}
\left. R \right|_{Z=0} = \frac{D}{D-2} \bar{R} \; .
\end{eqnarray}

As usual the asymptotic behaviour of the leading scalar fall off near the conformal boundary is $f_{\omega} \sim z^{\Delta - D}$ in these coordinates. Hence more generally this leading fall off can be written in terms of the 
function $Z$ as $f_{\omega} \sim Z^{\Delta - D}$.
This leading decay is set to zero by our Dirichlet condition. The free (normalisable) subleading behaviour is $f_{\omega} \sim z^\Delta$ 
in our normal coordinates, so generally may be written as $f_{\omega} \sim Z^\Delta$.

%
\subsection{The bulk geometry}
\label{sec:bulkgeom}
%

We have restricted the static bulk geometry to be smooth 
in its interior and have no asymptotic regions, boundaries or horizons other than the conformal boundary whose class is given by equation \eqref{eq:CFTmetric} with $(\Sigma, \bar{g})$ a closed space. This has important consequences. Before we discuss these it is useful to see how some examples work in detail.

The canonical example of such a static bulk spacetime is global AdS itself, with $\Sigma$ being the round sphere with radius $\mathcal{R}$, so,
\begin{eqnarray}
\label{eq:globalAdS}
Z &=& \mathcal{R} \cos{\theta}  \nonumber \\
g_{ij} dx^i dx^j &=& \mathcal{R}^2 \left( d\theta^2 + \sin^2{\theta} d\Omega^2_{(D-1)} \right) = \mathcal{R}^2 d\Omega^2_{(D)}
\end{eqnarray}
where $\theta \in [0,  \frac{\pi}{2} ]$ 
and $d\Omega^2_{(n)}$ is the line element on the unit round $n$-sphere. 
Note that the optical metric is a round hemisphere, and hence has constant 
Ricci scalar $R = D(D-1)/\mathcal{R}^2$ - this will play an important role in what follows. 
The conformal boundary corresponds to the equatorial boundary of the hemisphere, $\theta = \pi/2$,
yielding
\begin{eqnarray}
ds_{CFT}^2 = -dt^2 + \mathcal{R}^2 d\Omega^2_{(D-1)} 
\end{eqnarray}
so that $\Sigma$ is indeed a round $(D-1)$-sphere radius $\mathcal{R}$. We note that the Ricci scalar of the boundary space $(\Sigma, \bar{g})$ is $\bar{R} = (D-1)(D-2)/\mathcal{R}^2$, which obviously agrees with~(\ref{eq:Rbb}) above as it must.
While global AdS falls in our class of bulk vacuum geometries,
both Poincare AdS, with its Minkowski boundary, and the hyperbolic slicing of AdS with its boundary $-dt^2 + \mathcal{R}^2 dH^2_{(D-1)}$, with $dH^2_{(D-1)}$ the line element on unit radius hyperbolic space,
 fail to fall within our conditions. Both contain bulk horizons and $\Sigma$ is not compact. While one may take discrete quotients to yield compact $\Sigma$ in both cases, the bulk will still not obey our constraints.

Another example is the (toroidal identification of the) AdS-soliton \cite{Horowitz:1998ha}, where the boundary $\Sigma$ is $\mathbb{T}^{D-2} \times S^1$, where the periods of the torus are arbitrary, and the $S^1$ has length $L$ and the CFT is chosen to have Scherk-Schwarz boundary conditions about it. Then the appropriate bulk metric dual to the CFT vacuum is,
\begin{eqnarray}
Z = z \; , \quad g_{ij} dx^i dx^j = \left( 1 - \left( \frac{z}{z_0} \right)^D \right)^{-1} dz^2 + ds^2_{T^{D-2}} + \left( 1 - \left( \frac{z}{z_0} \right)^D \right) d\theta^2 
\end{eqnarray}
where $ds^2_{T^{D-2}}$ is a flat metric on the $(D-2)$-torus, and the parameter $z_0$ is determined by $L$ as $z_0 = D \, L / 2$ so that the $S^1$ closes smoothly at $z = z_0$ in the bulk. In this case we may compute the optical Ricci scalar to find,
\begin{eqnarray}
R = D (D-1)\frac{z^{D-2}}{ z_0^D } \; .
\end{eqnarray}
This vanishes at the boundary, which again is consistent with~(\ref{eq:Rbb}), since the spatial boundary metric is simply $ds^2_{T^{D-2}} + d\theta^2$ and thus Ricci flat. Since the boundary is Ricci flat, this example will be of less interest to us even though it satisfies our conditions on the bulk. We will primarily be concerned with $(\Sigma, \bar{g})$ with positive scalar curvature. We emphasise however, that it is obvious the AdS-soliton may be deformed by taking $ds^2_{T^{D-2}} \to d\sigma^2_{(D-2)}$, where $d\sigma^2$ is a closed Einstein space whose curvature is small compared to the size of the $\theta$ circle.\footnote{
This is because the Einstein equations do not depend on the topology of $ds^2_{T^{D-2}}$, but only the fact it is an Einstein space. Thus any Einstein space may be substituted, and providing its curvature is very small compared to the curvature scale associated to the $\theta$ circle closing, a smooth solution to the ordinary differential equations for the metric components as a function of $z$ should still exist. We note a metric function must be included for the $d\sigma^2$ component, as well as the functions for $dz^2$ and $d\theta^2$. We have verified these solutions exist numerically.}
The bulk will exhibit the same energy gap behaviour. However now the boundary metric may have scalar curvature which is either positive (for example, $d\sigma^2$ is a round sphere) or negative (for example $d\sigma^2$ is a compact hyperbolic metric). This allows non-trivial examples of bulk spacetimes satisfying our conditions (no bulk horizons, and no boundaries or asymptotic regions other than the locally AdS one) with both positive but also negative scalar curvature, and topologies different from $\Sigma$ being a sphere. In particular, in the case of $D=4$ bulk geometries with $\Sigma = S^2 \times S^1$ with the $S^2$ having a round metric have been numerically constructed, and for a large $S^2$ compared to the $S^1$ the geometry was indeed found to be close to that of the AdS-soliton \cite{Copsey:2006br}.

Now let us consider the implications of our constraints on the bulk. Recall that the optical Ricci scalar obeys,
\begin{eqnarray}
\label{eq:Riccieqn}
R = - \frac{(D-1)}{Z} \nabla^2 Z  =  \frac{D(D-1)}{Z^2} \left( 1 - \left( \partial_i Z \right)^2 \right)
\end{eqnarray}
and away from the conformal boundary we have assumed $Z > 0$ with no other asymptotic regions or boundaries.
These assumptions imply $Z$ must have a finite positive maximum value somewhere. Let us call this finite value $Z_{max} > 0$. 
The fact that $Z$ attains a maximal value $Z_{max}$ at some point in the bulk implies that redshift is maximised there, so $\xi_{max} = Z_{max} / \ell$ at that point.
Physically this is a reflection of the fact that the bulk geometry leads to an energy gap. 

Let us denote the points in the base where the function $Z$ is extremised as $p_{(n)}$, so $0 = \partial_i Z |_{p_{(n)}}$. We denote the value of $Z$ and the optical Ricci scalar $R$ at these points as $Z_{(n)} = Z(p_{(n)})$ and $R_{(n)} = R(p_{(n)})$ respectively. 
Let us also denote the values of the redshift function $\xi_{(n)} = \xi(p_{(n)})$.
There must be at least one such point, say $p_{(1)}$, corresponding to the global maximum of $Z$, but generally there may be other extremal points.
Since from equation \eqref{eq:Riccieqn} we see $Z^2 R$ is maximised at extremal points of redshift, $\partial_i Z = 0$, then $R$ must also be extremised there too. Hence extremal points of $Z$ are extrema of $R$. 
At these extremal points we see from equations~\eqref{eq:Riccieqn} that,
\begin{eqnarray}
\label{eq:extremal}
D(D-1) = Z^2_{(n)} R_{(n)} \; , \quad \left.  \nabla^2 Z  \right|_{p_{(n)}} =  - \frac{D}{Z_{(n)}} < 0
\end{eqnarray}
Hence the extremal redshift values $Z_{(n)}$ are precisely related to the optical Ricci scalar at these points, $R_{(n)}$, and furthermore $Z$ cannot be a minimum at the points (i.e.  only a maximum or saddle point).
While extrema of $Z$ correspond to extrema of $R$ we have no converse argument  saying an extremal point of $R$ always coincides with an extremum of $Z$.

The fact that the optical geometry is smooth implies there is both a maximum and minimum value of the optical Ricci scalar. Let us denotes these as $R_{max}$ and $R_{min}$ respectively. We know $R_{max} > 0$ since $R_{(n)} > 0$.
Further, recall that since the boundary Ricci scalar $\bar{R}$ and the bulk optical Ricci scalar evaluated at the boundary, $R |_{Z=0}$, are simply related as in equation~(\ref{eq:Rbb}) then we have,
\begin{eqnarray}
R_{min} \le \frac{D}{D-2} \bar{R}_{min} \; , \qquad R_{max} \ge \frac{D}{D-2} \bar{R}_{max}
\end{eqnarray}
where $\bar{R}_{min}$ and $\bar{R}_{max}$ are the minimum and maximum values of the Ricci scalar of $\Sigma$, the spatial part of the geometry the CFT is defined on.

An important consequence of the bulk Einstein equations is a minimum principle for the optical Ricci scalar. It is a straightforward task to confirm that the bulk equations imply the optical Ricci scalar obeys the elliptic equation,
\begin{eqnarray}
\nabla^2 R - \frac{(D-3)}{Z} \partial^i Z \partial_i R = \frac{2}{D-1} \left( R^2 - D R_{ij} R^{ij} \right) \; .
\end{eqnarray}
For a Riemannian metric we have the inequality $R_{ij} R^{ij} \ge \frac{1}{D} R^2$, which yields, 
\begin{eqnarray}
\nabla^2 R - \frac{(D-3)}{Z} \partial^i Z \partial_i R \le 0
\end{eqnarray}
and hence an extremum of $R$ where $\partial_i R = 0$ must have, $\nabla^2 R \le 0$, and hence cannot be a minimum. This has the crucial implication that the optical Ricci scalar $R$ must be minimised at the conformal boundary, as it cannot be minimised within the  bulk, and by assumption there are no other asymptotic regions or boundaries.
Due to the simple relation of the boundary Ricci scalar and the bulk Ricci scalar this then implies,
\begin{eqnarray}
\label{eq:riccireln}
R_{min} = \frac{D}{D-2} \bar{R}_{min}
\end{eqnarray}
Thus knowing the boundary spatial metric $(\Sigma, \bar{g})$, and hence $\bar{R}_{min}$, actually bounds the bulk Ricci scalar from below. This will play an essential role in what follows.

We note that we cannot in general expect such control on $R_{max}$ as already the AdS-soliton is an example with vanishing boundary $\bar{R}$, but positive bulk $R$. It remains a possibility that requiring a spherical topology on the boundary one might hope to control the maximum bulk $R$ in terms of the boundary geometry, but we have not been able to do so to date.

%
\section{Bulk  scalar fluctuations}
\label{sec:holoscalar}
%

Given that the boundary spatial geometry $(\Sigma, \bar{g})$ is closed we expect our holographic CFT on $\mathbb{R}_t \times \Sigma$ to exhibit an energy gap and discrete spectrum. Let us now see how this arises from the bulk perspective. For simplicity we will consider dynamics in the bulk associated to small fluctuations of a minimally coupled scalar field. As discussed earlier, for a static vacuum of the CFT described by the dual universal $(D+1)$-dimensional gravity sector, any small fluctuation associated to a CFT scalar operator will be described by a minimally coupled bulk scalar field on this vacuum geometry. The mass of this scalar will depend on the origin of the mode and would have to be examined in a top-down model. We reiterate that we will consider masses such that the CFT scaling dimension of the dual scalar operator obeys $\Delta > D/2$. Our aim is then to use our assumptions on the bulk to show that the spectrum of fluctuations for such a bulk scalar which are determined from equation~\eqref{eq:modeeqn} is discrete and to bound the lowest lying mode in terms of CFT data in analogy with equation~\eqref{eq:freebound2} for the free field example.

For a bulk spacetime that ends in an extremal horizon, such as for example for Poincare AdS, the mode solutions of a scalar will not be normal modes with real 
$\omega^2$ as energy can escape through this horizon. However, given our assumptions on the bulk geometry, the eigenvalue problem~(\ref{eq:modeeqn}) is of self-adjoint form so we do indeed expect a discrete normal mode spectrum with real eigenvalues $\omega^2$. Let us briefly recall this standard argument.
Consider the eigenfunctions $f_{(a)}$ with eigenvalues $\omega^2_{(a)}$. We begin by writing,
\begin{eqnarray}
 \int_{\mathcal{M}} \sqrt{g} \left( \frac{1}{Z^{D-1}} \partial_i f_{(a)}^\star \partial^i f_{(b)} + \frac{\ell^2 m^2}{Z^{D+1}} f_{(a)}^\star f_{(b)} \right)  &=& \int_{\mathcal{M}} \sqrt{g} f_{(b)} \left( - \nabla^i \left( \frac{1}{Z^{D-1}} \partial_i f_{(a)}^\star \right) + \frac{\ell^2 m^2}{Z^{D+1}} f_{(a)}^\star \right) \nonumber \\
 &=& ({\omega^2_{(a)}})^\star \int_{\mathcal{M}} \sqrt{g} \frac{1}{Z^{D-1}} f_{(a)}^\star f_{(b)} 
\end{eqnarray}
and then rewrite the same expression by integrating by parts the other way,
\begin{eqnarray}
 \int_{\mathcal{M}} \sqrt{g} \left( \frac{1}{Z^{D-1}} \partial_i f_{(a)}^\star \partial^i f_{(b)} + \frac{\ell^2 m^2}{Z^{D+1}} f_{(a)}^\star f_{(b)} \right)  &=& \int_{\mathcal{M}} \sqrt{g} f_{(a)}^\star \left( - \nabla^i \left( \frac{1}{Z^{D-1}} \partial_i f_{(b)} \right) + \frac{\ell^2 m^2}{Z^{D+1}} f_{(b)} \right) \nonumber \\
 &=& \omega^2_{(b)} \int_{\mathcal{M}} \sqrt{g} \frac{1}{Z^{D-1}} f_{(a)}^\star f_{(b)} \; .
\end{eqnarray}
It is important that $\mathcal{M}$ is smooth, with finite volume, and with no boundary other than $\Sigma = \partial \mathcal{M}$ corresponding to the conformal one where 
$f_{(a)} \sim Z^{\Delta}$. 
Together with our condition $\Delta > D/2$, and assuming the eigenfunctions $f_{(a)}$ are bounded, these facts ensure the above integrals are convergent and that surface terms from integrating by parts vanish. 
The standard arguments then proceed by firstly considering $a=b$, so we deduce the eigenvalues $\omega_{(a)}^2$ are real
so we may choose real eigenfunctions too.
Then for non-degenerate eigenfunctions, so $\omega_{(a)} \ne \omega_{(b)}$, the modes are orthogonal so $\int_{\mathcal{M}} \sqrt{g} f_{(a)} f_{(b)} / Z^{D-1} = 0$. We may now prove the spectrum is discrete by contradiction. Suppose there is a continuous spectrum. Consider a one parameter subset of non-degenerate modes, 
$f_\omega = f(\alpha)$,
 with $\omega^2 =\omega^2(\alpha)$ where $\alpha$ is a continuous parameter. Then consider the orthogonality condition,
\begin{eqnarray}
0 = \left( \omega^2(\alpha)  - \omega^2(\alpha_0) \right)  \int_{\mathcal{M}} \sqrt{g} \frac{1}{Z^{D-1}} f(\alpha_0) f(\alpha) \end{eqnarray}
and differentiating with respect to $\alpha$ and setting $\alpha = \alpha_0$ implies,
\begin{eqnarray}
0 = \left. \frac{d \omega^2}{d\alpha} \right|_{\alpha_0}  \int_{\mathcal{M}} \sqrt{g} \frac{1}{Z^{D-1}} f(\alpha_0)^2 
\end{eqnarray}
and hence $d \omega^2/ d\alpha = 0$ at $\alpha_0$. Since this holds for any value $\alpha_0$, this implies $\omega^2(\alpha) = $constant, in contradiction with our original assumption of non-degeneracy. 
Had we broken our conditions on the bulk, for example as for Poincare AdS where there is a bulk horizon, and hence another boundary in the integrals, we would not have derived these standard results.

A simple argument shows that in the case of a non-negative mass the quantity $Z_{max}$ gives a lower bound on the energy gap, $\omega_{min}$,
where as for the free conformal scalar, the energy gap is $\omega_{min} = \min_{\{\omega\}} | \omega |$.
Consider the following relation derived by integrating the scalar equation for a single mode;
\begin{eqnarray}
\label{eq:simplefunctional}
\omega^2 = \frac{ \int_{\mathcal{M}} \sqrt{g} \left( \frac{1}{Z^{D-1}} \left( \partial_i f_\omega \right)^2 + \frac{\ell^2 m^2}{Z^{D+1}} f^2_\omega \right)  }{ \int_{\mathcal{M}} \sqrt{g} \frac{1}{Z^{D-1}} f^2_\omega } \; .
\end{eqnarray}
Again these integrals are convergent, and boundary terms generated by integration by parts vanish, given our assumptions.
Ignoring the positive first term in the numerator, which we note cannot vanish as $\partial f_\omega \ne 0$, and noting $Z_{max} \ge Z > 0$ in the interior of the bulk, implies the bound,
\begin{eqnarray}
\label{eq:scalarbound1}
\omega^2_{min} & > & \frac{\ell^2 m^2}{Z_{max}^2}   \; , \qquad m^2 \ge 0 \; .
\end{eqnarray}
However while this is a lower bound on the energy gap (for a non-negative mass) we are looking for a bound that constrains the gap purely in terms of the CFT data, $\Delta$ and $(\Sigma, \bar{g})$. While our assumptions on the bulk geometry guarantee a maximum of $Z$, we do not want a bound involving the actual value unless we can determine this from the CFT data.

%
\subsection{Lower bound on the energy gap}
%

We now show that the energy gap for bulk scalar fluctuations with $\Delta > D/2$ in fact obeys a bound that is precisely analogous to that for the free conformal scalar. Given a wave function $f_\omega$ in equation~(\ref{eq:modeeqn}) define a rescaled wavefunction function $J = f_\omega / Z^\alpha$, for some $\alpha$, then using the bulk equations of motion we find,
\begin{eqnarray}
0 &=& \nabla^2 J + \frac{ 2 \alpha - D + 1}{Z} \partial^i Z \partial_i J   + \left( \omega^2 - \frac{\ell^2 m^2 - \alpha \left( \alpha - D \right)}{Z^2} - \frac{\alpha^2}{D(D-1)} R \right)  J 
\; .
\end{eqnarray}
Now multiplying by $J Z^\beta$ with $\beta = 2 \alpha + 1 - D$ and integrating, and then integrating by parts (the surface terms vanish), one obtains after a cancellation of terms,
\begin{eqnarray}
\label{eq:omegaintermediate}
&& \omega^2 \int_{\mathcal{M}}  \sqrt{g} Z^\beta J^2   = \int_{\mathcal{M}}  \sqrt{g} Z^\beta \left( \left( \partial J \right)^2 + \frac{\alpha^2}{D(D-1)} R J^2 + \frac{ \ell^2 m^2 - \alpha \left( \alpha - D \right) }{Z^2} J^2\right) 
\end{eqnarray}
and hence we find the lower bound on $\omega^2$,
\begin{eqnarray}
\omega^2 & \ge & \frac{ \int_{\mathcal{M}}  \sqrt{g} Z^\beta J^2 \left(  \frac{\alpha^2}{D(D-1)} R  + \frac{ \ell^2 m^2 - \alpha \left( \alpha - D \right) }{Z^2} \right) }{ \int_{\mathcal{M}}  \sqrt{g} Z^\beta J^2 } \; .
\end{eqnarray}
For $\Delta > D/2$ all the above integrals are convergent, and surface terms in the derivation vanish.

In the case of positive $m^2 \ge 0$, and taking $\alpha = 0$ one may simply recover our previous bound~(\ref{eq:scalarbound1}).
More generally if we choose $\alpha$ such that $\ell^2 m^2 > \alpha (\alpha - D)$ then we obtain a bound on the energy gap, $\omega_{min}$,
\begin{eqnarray}
\omega^2_{min} & \ge & \frac{\alpha^2  {R}_{min}}{D (D-1)}  + \frac{ \ell^2 m^2 - \alpha \left( \alpha - D \right) }{Z_{max}^2} 
 \end{eqnarray}
 This bound is saturated when both the optical Ricci scalar $R$ and the function $J$ are constant.
We now recall that we may write the minimum optical Ricci scalar in terms of the CFT data $\bar{R}_{min}$ as in equation~\eqref{eq:riccireln} using our minimum principle.
Making the choice $\alpha = \Delta$ we may eliminate the term involving 
the unknown
$Z_{max}$, retaining only the term involving $R_{min}$ which we know in terms of CFT data. Thus we find,
\begin{eqnarray}
\label{eq:omegalower2}
 \omega^2_{min}   \ge  \frac{\Delta^2 \bar{R}_{min} }{(D-1)(D-2)}  
\end{eqnarray}
which using the same notation as for our free 
conformal scalar
example, $\Omega = \omega_{min}  \mathcal{R}$ and $\bar{R}_{min} = (D-1)(D-2)/\mathcal{R}^2$, then gives,
\begin{eqnarray}
\label{eq:omegalower3}
\Omega \ge \Delta \; .
\end{eqnarray}
Thus we arrive at precisely the same bound as in the free scalar case. As discussed earlier, this bound implies the normalised gap $\Omega$ is minimised for $(\Sigma,\bar{g})$ being a round sphere, as for a sphere $\Omega = \Delta$ is then determined by global conformal symmetry. The bulk is then global AdS, where indeed the minimum energy eigenmode for a scalar fluctuation has $f_\omega = Z^\Delta$ and $\omega^2 = \Delta^2/\mathcal{R}^2$.

%
\subsection{Upper bound on the energy gap}
%

For completeness we now consider an upper bound on the energy gap
derived using a standard variation argument.
The bound is not given in terms of solely CFT data and thus is intrinsically less interesting than the lower bound we have just given. However it is interesting as a similar form of bound will be seen later for small black hole thermodynamic behaviour. We begin from equation~(\ref{eq:omegaintermediate}) with the choice $\alpha = \Delta$ by defining the functional,
\begin{eqnarray}
I[J]  &=& \frac{   \int_{\mathcal{M}}  \sqrt{g} \, Z^{2 \Delta + 1 - D} \left( \left( \partial J \right)^2 + \frac{\Delta^2}{D(D-1)} R J^2  \right) }{ \int_{\mathcal{M}}  \sqrt{g} \, Z^{2 \Delta + 1 - D} J^2 }
\end{eqnarray}
Thus we now regard $f_\omega = Z^\Delta J$ as a trial wave function. 
We restrict $J$ to be a suitably smooth function on $\mathcal{M}$ (not necessarily one corresponding to an actual eigenfunction) 
so that the trial wave function has the correct fall off at the boundary. 
Now one may check that $\delta I = 0$ when $J = J_\omega + \delta J$ with $f_\omega = Z^\Delta J_\omega$ an eigenfunction, energy $| \omega |$, for any smooth $\delta J$, and in addition $I[ J_{\omega} ] = \omega^2$. Hence any trial function $J$ gives an upper bound on the lowest energy, $\omega_{min}$, which is saturated only when the trial function coincides with the lowest eigenfunction.

Let us take a specific trial wave function corresponding to $J = $constant. Then this implies an upper bound on the lowest frequency in terms of $R_{max}$, so,
\begin{eqnarray}
\label{eq:omegaupper}
{
\omega^2_{min} \le  \frac{\Delta^2 R_{max}}{D(D-1)}
}
\end{eqnarray}
Recalling for global AdS that our trial function precisely gives the lowest energy eigenfunction, we immediately see that this bound is sharp and saturated by global AdS. 
It is interesting to note that with our previous bound~\eqref{eq:omegalower2} we then have $R_{min} \le \frac{D (D-1)}{\Delta^2} \omega^2_{min} \le R_{max}$ with saturation for global AdS where the optical Ricci scalar is constant. Thus the optical Ricci scalar serves as an important constraint on the behaviour of the energy gap for scalar fluctuations.

%
\section{Static bulk black holes}
\label{sec:holobh}
%

From general arguments of the fluid-gravity correspondence \cite{Baier:2007ix,Bhattacharyya:2008jc} we expect  `large' static bulk black holes to exist whose energy and entropy density dependence on temperature is universal at high temperature scales compared to the characteristic curvature scales of the boundary spatial geometry $(\Sigma, \bar{g})$. However, given our assumptions on the properties of the bulk vacuum geometry we also expect `small' static black holes (small in the sense of Hawking-Page) whose existence is intimately related to $Z$ having a maximal value. Consider a timelike curve in the spacetime, with spatial position $x^i(\tau)$ as a function of proper time $\tau$. The redshift function $Z$ acts as a potential for such curves. In particular the bulk geodesic equation reduces to an equation on the spatial optical geometry $(\mathcal{M}, g)$,
\begin{eqnarray}
\frac{d^2 x^i}{d \tau^2} + \Gamma^{i}_{~~jk} \frac{d x^j}{d \tau} \frac{d x^k}{d \tau} = \frac{1}{2 \ell^2}\partial^i \left( Z^2 \right) + \frac{2}{Z} \partial_j Z \frac{dx^j}{d\tau} \frac{dx^i}{d\tau}
\end{eqnarray}
We therefore see that a particle may be static, so $dx^i / d\tau = 0$, at an extremal point $p_{(n)}$ of $Z$. Furthermore, only the local maxima of $Z$ are stable if the geodesic is perturbed. 
The saddles are unstable (and there are no minima).

Let us place a very tiny black hole at such a point, $p_{(n)}$, corresponding to a maximum of $Z$. By tiny we mean that the radius of the black hole is far smaller that the local curvature scales at that point $p_{(n)}$, and so in particular the radius is smaller than $\ell$. However, the black hole must still be large compared to the bulk Planck scale so that we may treat it semiclassically within gravity.
The black hole should remain stably at the point and will locally approximate a $(D+1)$-Schwarzschild solution, and hence have very small energy and high temperature (due to its negative specific heat). In all these senses it will be an analog of the small black hole in global AdS.
One could presumably treat this static black hole solution carefully using matched asymptotic expansions, but we will 
make the reasonable assumption that there is no obstruction to existence.

The value of redshift at the point $p_{(n)}$, given by $ \xi_{(n)} = Z_{(n)}/\ell$, then determines the physical characteristics of the small black hole as viewed from the boundary CFT. While the redshift of the point $p_{(n)}$ relative to the conformal boundary diverges, since $Z \to 0$ there, recall that the time measured by the CFT is (in our conformal frame) just the extension of the bulk time $t$ to the boundary. Hence $\xi_{(n)}$ provides the finite, physical redshift as measured by the CFT. 
Let our tiny black hole sitting at the point $p_{(n)}$ locally approximate a Schwarzschild black hole with radius $r_h$. 
Its area will just be given by the usual local expression,
\begin{eqnarray}
A = \Omega_{(D-1)} r_h^{D-1}
\end{eqnarray}
in the limit $r_h \to 0$.
At the semiclassical level this gives rise to an entropy $S_{CFT} = A/4 G_{(D+1)}$ which will be the entropy associated to the small black hole in the CFT, which appears as an approximately thermal state.
However, its local temperature or mass will be scaled by the relative redshift. Locally it will appear to have the usual Schwarzschild mass,
\begin{eqnarray}
16 \pi G_{(D+1)} M_{local} = (D-1) \Omega_{(D-1)} r_h^{D-2}
\end{eqnarray}
in the limit $r_h \to 0$.
However the energy measured in the CFT, $E_{CFT}$, will be redshifted so $E_{CFT} = M_{local} /  \xi(p_{(n)})$. 
By $E_{CFT}$ we mean the energy of the bulk containing the tiny black hole minus the energy of the vacuum bulk, so the energy `due to the presence of the tiny black hole'. (Note that the energy of the vacuum state will not generally vanish due to the curved space Casimir energy).
Hence we expect the thermodynamic relation for a very small static black hole associated to the point $p_{(n)}$ in the vacuum geometry to be,
\begin{eqnarray}
\frac{E_{CFT}}{c_{}} & = & \frac{(D-1) \Omega_{(D-1)}}{\ell \, \xi_{(n)}} \left( \frac{ S_{CFT} }{4 \pi \Omega_{(D-1)} c_{} }\right)^{\frac{D-2}{D-1}} 
\end{eqnarray}
where we recall $c = \ell^{D-1} / 16 \pi G_{(D+1)}$ is the CFT central charge.
For all static black holes, including these very small ones, we expect $E_{CFT}, S_{CFT} \sim O(c)$ in the limit of large CFT degrees of freedom where a dual gravitational description will be valid. Thus it is convenient to define the energy and entropy densities normalised by the CFT central charge, $\tilde{E} = E_{CFT} / c$, $\tilde{S} = S_{CFT}/c$. In particular $\tilde{S} = 4 \pi A / \ell^{D-1}$ and hence measures the black hole size relative to the AdS length scale $\ell$. For these tiny black holes we should have $\tilde{S} \ll 1$, but still $O(1)$ in the gravity limit $c \to \infty$. We may  conveniently write the above relation for very small static black holes associated to the point $p_{(n)}$ as,
\begin{eqnarray}
\label{eq:smallbh}
\tilde{E} & = & \frac{a_D}{\ell \, \xi_{(n)} } {\tilde{S}}^{\frac{D-2}{D-1}} 
\end{eqnarray}
with $a_D = (D-1) \Omega_{(D-1)}/\left( 4 \pi \Omega_{(D-1)} \right)^{\frac{D-2}{D-1}}$ a dimensionless constant depending only on bulk dimension.

As a check we can confirm the relation for 
global AdS-Schwarzschild in the limit of a very small black hole. Global AdS-Schwarzschild takes the form,
\begin{eqnarray}
ds^2 &=& -g( \rho ) dt'^2 + g(\rho)^{-1} d\rho^2 + \rho^2 d\Omega^2_{(D-1)} \nonumber \\
g(\rho) &=& \left( \frac{\rho}{\ell} \right)^2 + 1 - \left( 1 + \frac{\rho_h^2}{\ell^2} \right) \left( \frac{\rho_h}{\rho} \right)^{D-2}
\end{eqnarray}
where the horizon is at $\rho = \rho_h$. The horizon area is $A = \rho_h^{D-1} \Omega_{(D-1)}$, and this determines,
\begin{eqnarray}
\tilde{S} \simeq 4\pi \Omega_{(D-1)}  \left(\frac{\rho_h}{\ell}\right)^{D-1} 
\end{eqnarray}
Taking $\ell t = \mathcal{R} t'$, one can write the metric in Fefferman-Graham form,
\begin{eqnarray}
ds^2
& =& \frac{\ell^2}{z^2} \Big[ dz^2 -  \left( 1 + \frac{1}{2} \left( \frac{z}{\mathcal{R}} \right)^2 + \frac{1}{16} \left( \frac{z}{\mathcal{R}} \right)^4 - \frac{D-1}{D} \left(1 + \frac{\ell^2}{\rho_h^2} \right) \left( \frac{\rho_h}{\ell} \right)^D \left(\frac{z}{\mathcal{R}}  \right)^D + \ldots \right) dt^2 \nonumber \\
&&  \qquad \qquad 
+   \left( 1 - \frac{1}{2} \left( \frac{z}{\mathcal{R}} \right)^2 + \frac{1}{16} \left( \frac{z}{\mathcal{R}} \right)^4 + \frac{1}{D} \left(1 + \frac{\ell^2}{\rho_h^2} \right) \left( \frac{\rho_h}{\ell} \right)^D \left(\frac{z}{\mathcal{R}}  \right)^D + \ldots \right) \mathcal{R}^2 d\Omega^2
\Big] \nonumber \\
\end{eqnarray}
where the $\ldots$ also include terms in odd bulk dimension of order 
$O(z^D \log{z})$, associated to the conformal anomaly, which do not depend on $\rho_h$.
We then identify the boundary CFT metric as,
\begin{eqnarray}
ds^2 = -dt^2 + \mathcal{R}^2 d\Omega^2_{(D-1)}
\end{eqnarray}
and the stress tensor differenced from global AdS, $T^{sub}$, as,
\begin{eqnarray}
T^{sub} = \frac{c_{}}{\mathcal{R}^D}  \left(1 + \frac{\ell^2}{\rho_h^2} \right) \left( \frac{\rho_h}{\ell} \right)^D  \left( (D-1) dt^2 + \mathcal{R}^2 d\Omega^2 \right) \; .
\end{eqnarray}
This `subtracted' stress tensor measures the change in energy and pressure due to the presence of the black hole. Note that any conformal anomaly contribution to the stress tensor cancels in this subtracted stress tensor, since such terms depend only on the boundary metric and not the state the CFT is in.
The energy above the vacuum due to the presence of the black hole is found by integrating the energy density given by the $tt$ component of this subtracted stress tensor over the spatial boundary metric to give,
\begin{eqnarray}
\tilde{E} = \frac{D-1}{\mathcal{R}}   \left(1 + \frac{\ell^2}{\rho_h^2} \right) \left( \frac{\rho_h}{\ell} \right)^D  \Omega_{(D-1)} \; .
\end{eqnarray}
We note that in the large black hole limit, $\rho_h \gg \ell$ we may then deduce,
\begin{eqnarray}
\label{eq:hightemp}
\frac{\tilde{E}}{V} \simeq \left(D - 1\right) \left( \frac{\tilde{S}}{4 \pi V} \right)^{\frac{D}{D-1}}
\end{eqnarray}
where $V = \mathcal{R}^{D-1} \Omega_{(D-1)}$ is the spatial volume of the boundary. This is precisely the universal high temperature behaviour of large black holes, and is the same as the behaviour of the planar AdS-Schwarzschild black hole.
However, we are interested in the tiny limit, $\rho_h \ll \ell$ where we find,
\begin{eqnarray}
\label{eq:smallAdS}
\tilde{E} &\simeq & \frac{ (D-1) \Omega_{(D-1)}}{\mathcal{R}} \left( \frac{ \tilde{S}}{4 \pi \Omega_{(D-1)}} \right)^{\frac{D-2}{D-1}}
\end{eqnarray}
and hence,
\begin{eqnarray}
\label{eq:smallAdS2}
\tilde{E}   & \simeq &  \frac{a_D}{\mathcal{R} } \tilde{S}^{\frac{D-2}{D-1}} \; .
\end{eqnarray}
Now we recall our earlier discussion of global AdS where $Z$ is given in~\eqref{eq:globalAdS} so that $Z$ has only one extremum at the point $p_{(1)}$ where $Z$ is maximised with $Z_{(1)} = \mathcal{R}$. The redshift there is then $\xi_{(1)} = \mathcal{R}/\ell$ and so we see that the relation in equation~(\ref{eq:smallbh}) is indeed true for very small global AdS-Schwarzschild black holes.

In summary, the CFT contains approximately thermal states which at very high temperature relative to the size of spatial curvatures in $\Sigma$ have two behaviours. The high energy states are universal, with behaviour the same as for planar AdS-Schwarzschild, or the 
very large energy global AdS black holes, as in equation~(\ref{eq:hightemp}).
Such states exist independent of the nature of the bulk geometry and in particular whether there is an energy gap. 
Conversely the low energy states depend critically on the 
details of the bulk geometry supporting the energy gap, although in the low energy limit only through the values $Z_{(n)}$, 
as in equation~(\ref{eq:smallbh}) above.
As for global AdS, these different branches of solutions presumably are asymptotics of the same moduli space of static black hole solutions, although with multiple small black hole branches one could imagine a complicated set of mergers perhaps of the form found for Kaluza-Klein black holes \cite{Horowitz:2011cq} as the energy is increased.

%
\subsection{Bounds on small black holes}
%

Given the behaviour in equation~\eqref{eq:smallbh} we immediately find sharp bounds in terms of the optical Ricci scalar given that at an extremal point we have the relation between $Z_{(n)}$ and $R_{(n)}$ in~\eqref{eq:extremal}. Thus we obtain,
\begin{eqnarray}
\tilde{E}^2 {\tilde{S}}^{-\frac{2(D-2)}{D-1}} & = & \frac{a_D^2}{Z_{(n)}^2 } = \frac{a_D^2}{D (D-1)} R_{(n)}
\end{eqnarray}
and since the optical Ricci scalar is, by assumption, a smooth function
in the interior of $\mathcal{M}$
we deduce,
\begin{eqnarray}
\label{eq:smallbhboundcurv}
{R}_{min}  \le \frac{D(D-1)}{a_D^2} \tilde{E}^2 \tilde{S}^{-\frac{2(D-2)}{D-1}}   \le R_{max} \; .
\end{eqnarray}
Clearly these bounds are sharp and saturated for global AdS as then the optical Ricci scalar is constant, so that $R_{min} = R_{max}$. We note that the upper bound takes a similar form to the upper bound on our scalar fluctuation gap in equation~\eqref{eq:omegaupper}. Now applying our minimum principle and constraints on the bulk geometry so that we have the relation 
 in equation~\eqref{eq:riccireln} then we obtain a lower bound on the very small black hole thermodynamics,
 \begin{eqnarray}
\label{eq:smallbhbound}
\bar{R}_{min}  \le \frac{(D-1)(D-2)}{a_D^2}  \tilde{E}^2 \tilde{S}^{-\frac{2(D-2)}{D-1}}  
\end{eqnarray}
which is now a bound entirely determined by CFT data. This equation is the analog of the scalar fluctuation bound~\eqref{eq:omegalower2}. 
Since the dual CFT energy for the black hole normalised by central charge, $\tilde{E}$, 
is dimensionful it is convenient to rewrite this relation in terms of a dimensionless $\tilde{\epsilon} = \tilde{E} \mathcal{R}$, where as before $\mathcal{R}$ is the radius of the sphere whose scalar curvature matches $\bar{R}_{min}$. Then we obtain an analog of the scalar bound \eqref{eq:omegalower3} as, 
 \begin{eqnarray}
\label{eq:smallbhbound2}
a_D \tilde{S}^{\frac{(D-2)}{D-1}}   \le \tilde{\epsilon}  
\end{eqnarray}
where the quantities entering this bound are invariant under a global scaling of the CFT metric.

%
\section{Discussion}
\label{sec:discussion}
%

Let us briefly review our main results. We have investigated the consequences of placing a $D$-dimensional CFT on a static curved spacetime, $\mathbb{R}_t \times \Sigma$ for a closed smooth spatial geometry $(\Sigma, \bar{g})$.  For a free conformal scalar it is simple to show that the theory exhibits a gap which is constrained as in equation~\eqref{eq:omegalower3},
\begin{eqnarray}
\Omega \ge \Delta
\end{eqnarray}
where $\Omega = \omega_{min}  \mathcal{R}$ is the 'normalised gap', a scale invariant measure of the energy gap, and $\mathcal{R}$ is the radius of a sphere which has the same scalar curvature as the minimum scalar curvature, $\bar{R}_{min}$, of $(\Sigma, \bar{g})$. Here $\Delta$ is the dimension of the scalar primary field, which for a free scalar is simply $\Delta = (D-2)/2$. The result may be interpreted as saying this normalised gap is minimised for $\Sigma$ being a sphere, and then the value of the righthand side of the bound above follows from global conformal invariance.

Our main result is that for a holographic CFT placed on the same spacetime, in a vacuum state whose dual is consistently described by $(D+1)$-dimensional gravity - the `universal gravity sector' - then precisely the same bound applies for fluctuations associated to a scalar primary field with a dual gravitational description. In any top-down model fluctuations of a generic scalar CFT operator which admit a gravitational description will correspond to dynamics of a minimally coupled scalar field about this pure gravity vacuum solution. Under reasonable assumptions on the geometry of this vacuum - in particular the lack of bulk horizons, boundaries or other asymptotic regions - we have proved the above bound holds.

While the dependence of the bound on the minimum value of the Ricci scalar of $\Sigma$ emerges in an obvious fashion for the free conformal scalar, it is rather subtle how this emerges in the bulk description for the holographic CFT. In the free case there is a direct coupling of the field to the Ricci scalar $\bar{R}$. However for the holographic CFT the bulk field is minimal and has no direct coupling to curvatures at all, and yet the bound again does depend on this Ricci scalar. This raises the question whether the bound above applies more generally to other CFTs. The statement is a reasonable one - when considering the energy gap associated to scalar fluctuations about the vacuum state for a CFT on the ultrastatic spacetime $\mathbb{R}_t \times \Sigma$, then this gap normalised by the minimum scalar curvature of $\Sigma$ is minimised for $\Sigma$ being a sphere. 
To be more precise, by energy gap we mean the lowest Hamiltonian eigenvalue for a state with a one point function of the scalar that is non-trivial.
Note that for multiple scalar primaries we would expect a bound for each field. Certainly in the free and holographic contexts this is the case.
An obvious future direction for investigation is to see whether this result can be obtained from general CFT methods. 
Another question is whether analogous bounds apply in the free and holographic context for fluctuations of other types of operator, such as fermions, vectors and the stress tensor, and if so, whether these also might apply to more general CFTs. 

For holographic CFTs static bulk black holes describe the behaviour of massive states in the CFT whose energies $E \sim O(c)$. The most massive states will correspond to very large black holes in the bulk, whose behaviour is universal, as in~\eqref{eq:hightemp}, where there is no dependence on the choice of spacetime $(\Sigma, \bar{g})$ since the temperature scale is far higher that the scales associated to $\Sigma$. The lower mass states with energies $E \sim O(c)$ will correspond to small black holes. These do not have a universal behaviour, and do depend on the details of the spacetime the CFT is placed on. However we have argued that in the very small limit, the thermodynamics of these black holes - which maps to the density of states in the CFT - has a  dependence on $(\Sigma, \bar{g})$ which simplifies, and implies the states obey the bound~\eqref{eq:smallbhbound2},
 \begin{eqnarray}
 (D-1) \Omega_{(D-1)} \left( \frac{S_{CFT}}{4 \pi \Omega_{(D-1)} c } \right)^{\frac{(D-2)}{D-1}}   \le \frac{ E_{CFT} \mathcal{R} }{c} 
\end{eqnarray}
where $E_{CFT}$ and $S_{CFT}$ are the energy and entropy associated to the black hole states, and the scale dependence of $E_{CFT}$ is normalised by precisely  the same quantity $\mathcal{R}$ as in the scalar fluctuation bounds. As for the scalar fluctuations this bound is sharp and saturated for global AdS, so the case $\Sigma$ is a round sphere.
Perhaps the most interesting aspect of this is the reemergence of the minimum scalar curvature $\bar{R}_{min}$ as the data which  normalises quantities associated to an energy gap. 

We expect that the branches of static small black holes we have considered, each associated to an extremal point of the function $Z$ in the vacuum, grow as their energies are increased and connect to the large static black hole branch of solutions, presumably via horizon mergers. 
We also expect that for finite horizon area (and entropy), these solutions will not have vanishing temperature, and hence there will exist a minimum temperature static black hole. 
Note that for $D=3$ it is known that no static extremal bulk black holes exist with spherical or toroidal horizon topology  \cite{Kunduri:2013gce}.
The main physical quantity of interest is this minimum temperature a static black hole can carry. A naive extrapolation of the universal high energy behaviour, and our low energy black hole behaviour bounded as it is above, suggests that the minimum Ricci scalar $\bar{R}_{min}$ of the CFT spacetime $(\Sigma, \bar{g})$ may again bound the minimum temperature from below. Thus a very interesting future direction would be to explore whether for a holographic CFT the minimum black hole temperature, normalised again by $\mathcal{R}$, is minimised for $\Sigma$ being a sphere. Similarly we would like to know if the deconfinement phase transition temperature (which is different to this minimal temperature since it is a first order transition) obeys a similar bound.

Our bounds may potentially be of interest for holographic models of confinement. There are two paradigms for confinement in holography. Either one may start with a higher dimensional CFT, and compactify some spatial direction as in \cite{Witten:1998zw,Sakai:2004cn}, or alternatively one may have an internal space in the bulk that caps off \cite{Klebanov:2000hb}. It is the former case that may be described within the universal gravity sector. Confinement associated to compactifying on a circle as in \cite{Witten:1998zw,Sakai:2004cn} may be described with the AdS-soliton bulk. In this case, as there is no curvature associated to this circle, our bounds give no information about the gap induced.
However, if one were to consider a CFT compactified on a space of two or more dimensions down to Minkowski (which may trivially be made spatially compact as in our discussion of the AdS-soliton in section~\ref{sec:bulkgeom} so that it obeys the conditions of our arguments), then the original spacetime may have non-trivial scalar curvature, and the gap induced for the effective lower dimensional theory would obey the bounds we have described here.

%
\section*{Acknowledgements}
%

We would like to thank Jerome Gauntlett and Paul McFadden for useful comments. This work was supported by the STFC grant ST/J0003533/1. TW would like to thank the Banff International Research Station workshop 15w5148, {\it Geometric Flows: Recent Developments and Applications}, where some of this work was completed.
AH is supported by an STFC studentship.

%
\bibliographystyle{apsrev4-1}
\bibliography{paperV1}
%

\end{document}